\documentclass[preprint,12pt]{elsarticle}
\journal{Computer Physics Communications}
\usepackage[T1]{fontenc} 

\usepackage{url}
\usepackage{graphicx}
\usepackage{color}
\usepackage{framed}

\newcommand{\be}{\begin{equation}}
\newcommand{\ee}{\end{equation}}
\newcommand{\bea}{\begin{eqnarray}}
\newcommand{\eea}{\end{eqnarray}}

\newcommand{\ep}{\varepsilon}

\newenvironment{code}
  {\tt\noindent\begin{framed}\begin{flushleft}}
  {\end{flushleft}\end{framed}\normalfont}

\begin{document}

\begin{frontmatter}

\title{FIESTA5: numerical high-performance Feynman integral evaluation}

\author[SRCC,MC]{A.V.~Smirnov\corref{cor1}}
\ead{asmirnov80@gmail.com}

\author[CS,MC]{N. D. Shapurov}

\author[CS,MC]{L. I. Vysotsky}

\cortext[cor1]{Corresponding author}

\address[SRCC]{Research Computing Center, Moscow State University, Moscow, Russia}
\address[CS]{Faculty of Computational Mathematics and Cybernetics, Moscow State University, Moscow, Russia}
\address[MC]{Moscow Center for Fundamental and Applied Mathematics, Moscow, Russia}

\begin{abstract}
In this paper we present a new release of the FIESTA program (Feynman Integral Evaluation by a Sector decomposiTion Approach). FIESTA5 is performance-oriented --- we implemented improvements of various kinds in order to make Feynman integral evaluation faster. We plugged in two new integrators, the Quasi Monte Carlo and Tensor Train. At the same time the old code of FIESTA4 was upgraded to the C++17 standard and 
mostly rewritten without self-made structures such as hash tables. There are also several essential improvements which are most relevant for complex integrations --- the new release is capable of producing results where previously impossible.
\end{abstract}
\end{frontmatter}
\newpage

{\bf PROGRAM SUMMARY}

\vspace{1cm}

\begin{small}
\noindent
{\em Manuscript Title:} FIESTA5: numerical high-performance Feynman integral evaluation\\
{\em Authors:} A.V. Smirnov, N. D. Shapurov, L. I. Vysotsky\\
{\em Program title:} FIESTA5\\
{\em Licensing provisions:} GPLv3\\
{\em Programming language:} {\tt Wolfram Mathematica} 8.0 or higher, {\tt C++}\\
{\em Computer(s) for which the program has been designed:} from a desktop PC to a supercomputer\\
{\em Operating system(s) for which the program has been designed:} Unix, Linux, Mac OS X, Windows (under WSL)\\
{\em RAM required to execute with typical data:} depends on the complexity of the \\ problem  \\
{\em Has the code been vectorized or parallelized?:} both\\
{\em Number of processors used: } from 1 processor up to loading a supercomputer; from a personal GPU up to professional GPUs at a supercomputer \\
{\em Supplementary material:} The article, usage instructions in the program package, https://bitbucket.org/feynmanIntegrals/fiesta
{\em CPC Library Classification:} 4.4 Feynman diagrams, 4.12  Other Numerical
Methods, 5 Computer Algebra, 6.5 Software including Parallel Algorithms\\
{\em External routines/libraries used:} {\tt Wolfram Mathematica} [1], {\tt KyotoCabinet} [2], {\tt Cuba} [3], {\tt QMC} [4], {\tt Tensor Train} [5], {\tt mimalloc} [6]\\
{\em Nature of problem:}
Sector decomposition is a well-known approach to the numerical evaluation of Feynman integrals.
Feynman integrals in 4 space-time dimension are divergent and have to be regulated. Sector decomposition
is used to resolve pole singularities and consists of different stages --- sector decomposition itself, contour decomposition (in case of physical kinematics meaning base functions changing sign therefore leading to integration in complex numbers), pole resolution, epsilon expansion and numerical integration. 
\\
{\em Solution method:}
Most stages are performed in {\tt Wolfram Mathematica} [1] (required version is 8.0 or higher), this part
is parallelized by the use of {\tt Mathematica subkelnels} in shared memory. As a result a database on hard disk is produced with the use of the {\tt KyotoCabinet} [2] database engine. The integration stage is written in {\tt C++} and can be run on personal computers as well as on supercomputers via {\tt MPI}. It can make use of installed graphical processor units. As default integrator we use Vegas from the Cuba library [3], but also it can be switched to QMC [4] or Tensor Tran [5]. The mimalloc memory allocator [6] can be used for improved performance.
\\
{\em Restrictions:} The complexity of the problem is mostly restricted
by CPU time required to perform the integration and to obtain the desired precision.\\
{\em Running time:} depends on the complexity of the problem.\\
{\em References:} 
{\\} [1] http://www.wolfram.com/mathematica/, commercial algebraic software; 
{\\} [2] http://fallabs.com/kyotocabinet/, open source;
{\\} [3] http://www.feynarts.de/cuba/, open source; 
{\\} [4] https://github.com/mppmu/qmc/, open source; 
{\\} [5] https://bitbucket.org/vysotskylev/c-tt-library, open source; 
{\\} [6] https://github.com/microsoft/mimalloc.git, open source;

\end{small}

\newpage

\section{Introduction}

Sector decomposition is an already long-established method of numerical Feynman integral evaluation.
While sector decomposition itself is known from mathematician papers from previous century,
in particle physics was first introduced by Binoth and Hienrich~\cite{Binoth:2000ps,Binoth:2003ak,Binoth:2004jv,Heinrich:2008si} and the first public version of a sector decomposition program was published
by Bogner and Weinzierl~\cite{Bogner:2007cr,Bogner:2008ry}.

Nowadays there are two well-known public competitive programs for sector decomposition,
{\tt SecDec} by Binoth and Heinrich~\cite{Heinrich:2008si},
later improved and made public by other collaborators and at some point renamed to {\tt pySecDec}~\cite{Carter:2010hi,Borowka:2012yc,Borowka:2012ii,Borowka:2012rt,Borowka:2013cma,Borowka:2013lda,Borowka:2014koa,Borowka:2015mxa,Borowka:2017idc,Jahn:2018jml,heinrich2021expansion}
and {\tt FIESTA}~\cite{Smirnov:2008py,Smirnov:2009pb,Smirnov:2013eza,Smirnov:2015mct}.

In this paper we are not going to repeat detailed explanations what sector decomposition is to avoid 
text borrowing (there is no original way to present sector decomposition so many times)
and are going to focus on how the sector decomposition can be made effective taking most 
of CPUs and GPUs in use.

The new release of {\tt FIESTA} is a significant upgrade of the previous {\tt FIESTA} version.
The key features of the new release are
\begin{itemize}
 \item improvement of contour decomposition algorithms in physical kinematics making {\tt FIESTA} work in cases where it was previously impossible and much faster; 
 \item new integrators are available, the Quasi Monte Carlo algorithm~\cite{BOROWKA2019120} and Tensor Train algorithm~\cite{Vysotsky2021};
 \item multiple internal code optimizations including full support for AVX (single instruction multiple data) processor instructions resulting in a faster integrand evaluation;
 \item multiple options aimed at faster evaluation, for example, the possibility to seed a different number of sampling points in different sectors depending in their contribution to the final result (balancing);
 \item the integral evaluation was measured with code profilers and some parts of the code taking unreasonably large amounts of time were improved.
\end{itemize}

At the same time the new release is also a serious update of the codebase:
\begin{itemize}
 \item the {\tt Mathematica} part of the code is now a {\tt Mathematica} package without influencing global context,
 all functions have a usage explanation;
 \item legacy code with self-made hash tables, tries and vectors was removed and replaced with appropriate std structures;
 \item the code standard was upgraded from {\tt c99} to {\tt C++17};
 \item configure script was added making it possible to set up user preferences before compilation;
 \item the code was covered with doxygen documentation and partially with tests (any new change is automatically tested so that it does not break the behavior in many situations).
\end{itemize}

While the points listed above do not produce direct benefit for a general user, we pose them as an important advancement since now a larger team of researchers might support and develop the code,
and we have plans for future advancement and releases.

The rest of the paper is organized the following way: in the next section we give a brief description 
of the sector decomposition approach mainly focusing on stages of this method since their understanding 
is essential for setting proper options for {\tt FIESTA} optimization. We also present the new features
of {\tt FIESTA5} related to each of the stages. In section~\ref{installation} we explain how to install {\tt FIESTA} and how to use it. In section~\ref{options} we list all {\tt FIESTA} options and provide comments on how to set them properly depending on the example in use. In section~\ref{benchmarks} we provide benchmarks comparing the usage with different options and different versions of {\tt FIESTA}.

\section{Sector decomposition stages}

The sector decomposition approach is a complex method consisting of multiple stages, performed automatically one after another. Different sector decomposition programs might have somewhat different approaches, so we are going to list those stages in the way they are used in {\tt FIESTA} and give their characteristics and possible issues.

Sector decomposition starts from an integral over $n$ variables from $0$ to $\infty$ but this integration is actually in finite ranges since the integrand contains a delta-function of the sum of all variables that is a function that is equal to $1$ if the sum is equal to $1$ and $0$ otherwise. Apart from the delta function the integrand consists of a product of polynomials of integration variables with the exponent depending from a special variable, $\ep$ where $d = 4 - 2 \ep$ us the space-time dimension. This $\ep$ is small and we are interested in the first terms of the series of the integral in $\ep$.

For a Feynman integral the formula is the following:

\begin{eqnarray}\label{Alpha}
    &&\mathcal F(a_1,\ldots,a_n) =(i\pi^{d/2})^l\times
   \\\nonumber
    &&\frac{\Gamma(A-l d/2)}{\prod_{j=1}^n \Gamma(a_j)}
            \int_{x_j\geq 0} d x_i\ldots d x_{n} \delta\left(1-\sum_{i=1}^n x_i \right)
                \left(\prod_{j=1}^n x_j^{a_j-1}\right) \frac{U^{A-(l+1)d/2}}{(F-i0)^{A-ld/2}},                         
\end{eqnarray}

Here $l$ is the number of loops of the Feynman diagram, $A$ is the sum of indices and $U$ and $F$ are certain polynomial of integration variables that are constructively defined based on the diagram. We should also note that the approach is valid for more general integrals. We are going to list all stages we use, for details one can refer to previous papers on {\tt FIESTA}.

If F is negative then there is no imaginary part. In particular, this happens in regions where all the kinematic invariants are negative. For physical values of the kinematic invariants and the masses, this function can be positive so that a given Feynman integrals has a non-zero imaginary part.

\subsection{Dealing with negative indices}

If some of the indices are non-positive integers, the integration is performed according to the rule (here $f^{(n)}$ is the $a$-th derivative of $f$):
\begin{eqnarray}
\lim_{a\rightarrow -n} \int_{0}^{\infty}d x \frac{x^{(a-1)}}{\Gamma(a)} f(x) = f^{(n)} (0)
\label{negative}
\end{eqnarray}

\textit{Complexity}: low;\\\indent
\textit{Parallelization}: no;\\\indent
\textit{Possible issues}: the resulting expression becomes more complex, therefore if one is aiming at results with reasonable precision, integrals without negative indices should be chosen as master integrals (the integrals that one reduced all integrals for a chosen diagram to).

\subsection{Preresolution}

The problem of integrals of equation~\ref{Alpha} is that one should expand in $\ep$ before proceeding to numerical integration, but one cannot just change the order of integration and the expansion operations. To be able to do this one should first reveal possible singularities. While the basic sector decomposition approach aims at revealing singularities of the $x[i]^{-1+\ep}$ type, the singularities might be also inside the integration region due to hidden $(x[i]-x[j])^{-1+\ep}$ terms.

The preresolution stage as we call it aims at getting rid of the singularities of a special form. By default it is off in the complex mode, meaning physical kinematics where $F$ can change sign. But this stage is especially important in threshold configurations (which are defined by an equality between the square of the momentum flowing thorugh a cut of a given graph and the square of the sum of the masses in the threshold) where $F$ tends to zero not only due to $x[i] \rightarrow 0$ but also due to $x[i] \rightarrow x[j]$ and some more complex combinations. In this case we start searching for pairs of $i$ and $j$ such that dividing the integration region in two zones ($x[i] \leq x[j]$ and vice versa) and performing a change $x[i] \rightarrow x[i] + x[j], x[j] \rightarrow x[j]$ (and a similar one in the opposite region) decreases the number of negative terms in $F$. This stage can increase the number of starting sectors for sector decomposition.

\textit{Complexity}: low;\\\indent
\textit{Parallelization}: no;\\\indent
\textit{Related options}: {\tt NegativeTermsHandling};\\\indent
\textit{Possible issues}: none.

\subsection{Sector decomposition}
Sector decomposition itself is the basis of the approach. This paper is not pretending to be a detailed presentation of the method, so we are going to skip details, but to be short, it is performed after removing the delta function in equation~\ref{Alpha} (so the integration region is at this point a unitary hypercube) and an iterative sector decomposition step takes an expression of the $x[i] + x[j]$ type and again splits the integration region in two zones ($x[i] \leq x[j]$ and vice versa) and performs a change $x[i] \rightarrow x[i] * x[j], x[j] \rightarrow x[j]$ and a similar inverse one. After variable replacements we are again back to the unitary hypercube, and the function takes the form $x[i] * (1 + x[j])$. The goal of the method is to end with a number of sectors (each of those being a unitary hypercube after variable change) so that in each of the sectors the expression is split into a product of a ``well-behaving'' term (a positive polynomial containing a constant) and monomials of integration variables.

There are different sector decomposition strategies implemented in {\tt FIESTA}. In previous versions the default strategy was {\tt S} (for details see~\cite{Smirnov:2008py}), that is our original strategy on sector decomposition. The other two reasonable possibilities were {\tt X}, as we understood the original strategy from SecDec, and also {\tt KU}, that is the Kaneko-Ueda strategy~\cite{Kaneko:2009qx}.

Although for following stages it is most profitable to have as small number of sectors as possible, and although the KU strategy is supposed to result in the smallest number of sectors in most cases, we do not set it as default for two reasons. First, it requires the external program {\tt qhull} to be installed to be able to find a convex hull. More importantly at high dimensions in equation~\ref{Alpha} (around $8-10$ or more depending on the diagram) the strategy itself starts taking too much time and sector decomposition can become the hardest stage making the strategy choice ineffective.

Note: a recent paper of Borinsky~\cite{Borinsky:2020rqs} is suggesting a tropical approach to sector decomposition, but it is based on the same Kaneko-Ueda strategy. Thus no new strategy was implemented in {\tt FIESTA} to match the new approach. What we found much more valuable in that paper is the possibility to estimate the contribution of  different sectors in advance and thus to use a different number of sampling points for different sectors. Inspired by this idea we suggest our way to do a similar thing, but not depending on the sector decomposition strategy. We will explain details in the subsection devoted to the integration stage.

\textit{New in {\tt FIESTA5}}: we now provide the {\tt KUS} and {\tt S2} strategies (the second is default now) performing sector decomposition for each function in the product ($U$, $F$ and such) separately. For each sector corresponding to the first of the functions the algorithm performs a variable substitution for the next and follows with a sector decomposition there if needed. This is not optimal by the number of sectors but seriously more efficient in time for this stage at high dimensions.

\textit{Complexity}: depends on the number of positive indices and strategy, might be high;\\\indent
\textit{Parallelization}: up to the number of primary sectors, with the use of {\tt Mathematica} subkernels;\\\indent
\textit{Related options}: {\tt Strategy, FixSectors, NumberOfSubkernels, \\ PrimarySectorCoefficients, MixSectors};\\\indent
\textit{Possible issues}: the strategy has to be chosen wisely. While the Kaneko-Ueda strategy {\tt KU} seems to be optimal according to the number of sectors, it might take too much time for this stage with $8$ or more positive indices.

\subsection{Symmetries search}

\textit{New in {\tt FIESTA5}}: at this stage we detect equal sectors. To do that we bring the polynomials to canonical form as in the original code {\tt tsort} of Pak~\cite{Pak_2012}.

\textit{Complexity}: low;\\\indent
\textit{Parallelization}: none\\\indent
\textit{Related options}: {\tt SectorSymmetries};\\\indent
\textit{Possible issues}: this option is switched off in case of expansion by regions ({\tt SDExpandAsy} and similar) since it might incorrectly identify sectors.

\subsection{Hypercube bisection}

While the sector decomposition approach deals with singularities due to $x[i] \to 0$, in some cases, normally in physical kinematics this might happen also due to $x[i] \to 1$. While being an unexpected equality, this spoils completely the integration in rare situations. 

\textit{New in {\tt FIESTA5}}: The older versions had options related to bisection before sector decomposition. Obviously this is inefficient since different sectors might require bisection of different types. The following strategy appeared in {\tt FIESTA5} and if turned on starts searching for variables responsible for turning $F$ to zero when $x[i] \to 1$ in each sector separately. If such variables are found, a bisection of the corresponding axis is performed dividing the sector in two ($x[i] \leq p$ and $x[i] \geq p$ for some $p$) and making a variable replacement returning both parts to the range from $0$ to $1$ (here $1$ is mapped to $0$ and $p$ to $1$).

After this stage an additional sector decomposition might be performed.

\textit{Complexity}: low;\\\indent
\textit{Parallelization}: efficient with the use of {\tt Mathematica} subkernels;\\\indent
\textit{Related options}: {\tt SectorSplitting, NumberOfSubkernels}\\\indent
\textit{Possible issues}: the option is turned off by default. The reason is that it produces extra sectors making following stages more complex, but the fact that $F$ turns to $0$ when $x[i] \to 1$ does not nesseseraly mean a numerical singularity, so turning this option on might be a waist of time. Turn it on is case of numerical problems.

\subsection{Contour transformation}

Contour transformation is a special stage required only in physical kinematics. If the function $F$ changes sign in the integration region then it is obviously equal to zero somewhere inside on a contour of an unspecified shape. While we are able to deal with singularities at $x[i] \to 0$ (with the sector decomposition and pole resolution), at $x[i] \to 1$ (with hypercube bisection) and at $x[i] \to x[j]$ and similar (with preresolution), this is algebraically not possible for a general shape. Thus the only way to proceed is to avoid the contour by moving into the complex plane. The formula for contour transformation was originally suggested by Binoth and Heinrich and has the form 

\begin{equation}
x[i] \to x[i] * (1 - I * \lambda_i * (1 - x[i]) * D[F, x[i]]) 
\end{equation}

for all integration variables. Obviously this formula does not change integration boundaries, and the idea is that $F$ and the derivatives of $F$ do not turn to zero simultaneously for Feynman integrals. 

The bad feature about this formula is that it makes the integrand much more complex. Moreover one has to choose the $\lambda_i$ wisely so that on the one hand, the absolute value of the integrand after variable replacement is far enough from the origin, on the other hand, we should not hit an other branch of complex functions not to produce completely incorrect results. 

This ``lambda search'' is the part of the sector decomposition approach which does not have a simple solution. Thus, like in SecDec, {\tt FIESTA} tries to choose the lambdas, but sometimes the default settings are not optimal.

\textit{New in {\tt FIESTA5}}: this ``lambda search'' was improved greatly compared to the previous versions of {\tt FIESTA}. Partially the improvement is due to the optimization of the {\tt Mathematica} code --- the ``lambda search'' stages mostly consist of substituting multiple numerical values in some functions, and this could be made a few times faster with the use of the {\tt Compile} function. However the most serious improvement is the possibility to estimate beforehand which of the integration variables might be responsible for the sign change. Those, that are not, are not replaced by default with the contour transformation formula.

The contour transformation now has the following substages:
\begin{itemize}
 \item balancing; here the shifts added to different variables are estimated; if the shift is greater than $1$ meaning that the complex part will be greater that the real part, we multiply the shift for this variable by the inverse of the estimate of the maximum;
 \item choosing, where we get rid of variable shifts that are not supposed to change the sign of $F$;
 \item maximum searching, where we estimate the maximum of $\lambda$, so that the cubic terms are smaller that the linear terms in the expansion of the shift (the quadratic terms do not matter since they are real, not complex, and the fifth-order terms should be much smaller);
 \item best lambda search, where we take a few possible $\lambda$ from 0 (not inclusive) to the one from the previous step and choose the one so that the minimum of the norm of the function is maximal for this choice.
\end{itemize}

\textit{Complexity}: average, but the very fact of running contour transformation increases the complexity of following stages a lot;\\\indent 
\textit{Parallelization}: close to linear by {\tt Mathematica} subkernels;\\\indent
\textit{Related options}: {\tt ComplexMode, MinimizeContourTransformation, \\ ContourShiftCoefficient, ContourShiftIgnoreFail, LambdaSplit, \\ ContourShiftShape, FixedContourShift, LambdaIterations,\\ NumberOfSubkernels};\\\indent
\textit{Possible issues}: The optimal setting of options, especially such that {\tt ContourShiftShape} and {\tt ContourShiftCoefficient} might be in some cases a sort of a game with no obvious solution. {\tt FIESTA} has default values for those options set as seems to be appropriate for a large class of integrals, but if the integration has a poor convergence, perhaps these options should be tuned.

\subsection{Pole resolution}

At this stage all the singularities of sector expressions are to be of the form $x[i]^{a} * (1 + \ldots )^b$, 
where the second function never turns to zero, but $a$ can be a linear function of $\ep$ with a non-positive constant part. At this point it is still impossible to change the order of integration and $\ep$-expansion, so this is solved by adding and subtracting the first terms of the Taylor series of $(1 + \ldots )^b$ in $x[i]$. The remainder is known to be finite even being multiplied by $x[i]^{a}$, and for the terms of the Taylor series the integration can be taken out analytically. 

There are other approaches to pole resolutions based on integration by part relations, but our experience shows worse numerical behavior for those approaches.

Note: in some cases an extra regularization variable is needed --- a small variable that tends to zero. In case it is used, before this step there is another pole resolution stage (by the regularization variable) and then another expansion stage by the regularization variable similar to $\ep$ expansion. These stages are very similar and will not be discussed separately.

\textit{New in {\tt FIESTA5}}: nothing.

\textit{Complexity}: average;\\\indent
\textit{Parallelization}: close to linear by {\tt Mathematica} subkernels;\\\indent
\textit{Related options}: {\tt ResolutionMode, NumberOfSubkernels};\\\indent
\textit{Possible issues}: in case for some reasons the sector decomposition resulted in a pure negative power of an integration variable, there is a crash at this stage. Normally it means that a regularization variable should be used.

\subsection{Epsilon expansion}

This stage is more or less straightforward --- each expression is combined and expanded in $\ep$ up to the required order. We do not use the {\tt Expand} function of {\tt Mathematica}, but a faster approach with differentiation.

\textit{New in {\tt FIESTA5}}: nothing.

\textit{Complexity}: average;\\\indent
\textit{Parallelization}: close to linear by {\tt Mathematica} subkernels;\\\indent
\textit{Related options}: {\tt NumberOfSubkernels};\\\indent
\textit{Possible issues}: the expressions might grow in size significantly at this stage, please monitor the RAM usage.

\subsection{Expression generation}

This might be a technical stage but we need to mention it for {\tt FIESTA} users. At this point integrands are translated from a {\tt Mathematica} format into a special format suitable for the {\tt C++} parser. This step might take longer than expected of such a trivial operation.

\textit{New in {\tt FIESTA5}}: in previous versions we analyzed all integrals trying to check whether they are exactly equal to zero, meaning that the expression is a non-simplified algebraic expression equal to a constant zero. This was a round-way for an integration problem sometimes resulting in an undefined answer for totally zero integrals. Now this is no longer needed, and this check is off by default. Thus this part is now much faster now since substituting multiple values in a huge function might be slow in {\tt Mathematica}.

\textit{Complexity}: average;\\\indent
\textit{Parallelization}: close to linear by {\tt Mathematica} subkernels;\\\indent
\textit{Related options}: {\tt ZeroCheckCount, OptimizeIntegrationStrings, \\AnalyzeWorstPower, NumberOfSubkernels};\\\indent
\textit{Possible issues}: none we know.

\subsection{Integration}

Integration is the final and in general the most time-consuming stage of the sector decomposition approach. While with the basic setup the integration might be not the longest stage, the point is that the better numerical precision one wishes for results, the more sampling points need to be taken during integration, while previous stages do not depend on the planned precision at all (but can affect the integration precision due to an improper choice of $\lambda$). Of course everything is not that straightforward, for example the strategy {\tt KU} might result in the smallest number of sectors thus resulting in simpler integrands, but for a high number of positive indices it takes too long to use this strategy. So ultimately previous stages should only serve the goal of preparing expressions for integration as simple as possible.

Anyway the point is that as soon as one prepared the database with integrands, the resulting precision depends greatly on the number of sampling points used during integration, so apart from other options which should be set up in a proper way, the more sampling points are taken the better the result is. Hence this should be the most optimized part of a sector decomposition program, so in {\tt FIESTA} this part is written in {\tt C++} and has multiple optimizations and possibility to run it on a cluster or with the use of graphical processing units.

Integrating of a function over a unitary hypercube requires the following. First the expression that arrived from {\tt Mathematica} is parsed and translated into a form suitable for a fast evaluation. This parser originated in old versions of {\tt FIESTA} and is very fast, taking a negligible part of time. Then an important part is to use a proper \textit{integrator} that is a library that seeds sampling points and uses function values in those points as information needed to produce a result and error estimate. The old versions of {\tt FIESTA} relied on the Vegas algorithm from the Cuba library~\cite{Hahn200578}. 

The other important part is the ability to evaluate sampling points fast. In the end, after the choice of the integrator the time required to obtained a good precision is linear to the number of sampling points requested by the integrand.

On the other hand, a complex Feynman integral is represented as a sum of a huge number of sector integrals, around 10 thousand or more. Those integrands are completely independent from each other, so this task ideally suits for the {\tt MPI} approach on a supercomputer where each sector integral can be evaluated on a separate node.

\textit{New in {\tt FIESTA5}}: There are many changes in the integration part of {\tt FIESTA}. To start with, there was a huge update of the code-base, rewriting more that half of the code. The reason is that there were large pieces of legacy code in pure {\tt c} with self-made structures that exist in modern {\tt C++}. And some structures were replaced with ones more suitable.

More importantly for the user of {\tt FIESTA} is that we now plugged in two additional integrators, the Quasi Monte Carlo (QMC) and Tensor Train (TT). The first of those, QMC~\cite{BOROWKA2019120}, is an integrator already used in {\tt SecDec} and suggested for use in {\tt FIESTA} by Heinrich in her talk at the KIT TTP software seminar. This integrator was modified in order to be able to request batches of sampling points, and we made a pull request to the original repository. The other integrator, TT~\cite{Vysotsky2021} is an original one never previously used in practical situations up to our knowledge.

The Tensor Train approach may be viewed as enhanced cubature-rule integration. The latter suffers from the ``curse of dimensionality'', i.e. its complexity grows exponentially with the number of variables. However, if the integrand is approximated by a function allowing low-parametric parametrization, the cubature sum can be computed efficiently. Tensor Train is one of tensor factorizations that may be used in this approach.

There are multiple internal optimizations removing extra memory allocations. Also we suggest to use the plugged in mimalloc memory allocator for performance. However one of the most important upgrades is the possibility to use AVX instructions. Already in previous versions of {\tt FIESTA} we used batches of points generated by integrators in order to request multiple sampling points at a time. This lets processors optimize caches for improved performance. 

Moreover this fits well the function evaluation scheme in {\tt FIESTA}. Unlike {\tt SecDec} we do not compile integrands but use our own parser to provide a fast evaluation method running by tetrads (operation, first operand address, second operand address, result address). Now if those operands are not just single values but arrays of those operands, the operations can be performed with the use of processor instructions that are capable of, for example, multiplying not just a pair of doubles but four pairs of doubles at a time. 

Partially compilers are capable to carry out optimizations of this sort for the current architecture, but it does not work with AVX. Therefore we implemented a possibility to directly use those instructions with the use of Intel intrinsic operations.

Another improvement is related to sector balancing. The results in different sectors might differ in size by a few orders in magnitude, therefore there is no sense to seed the same number of sampling points in all sectors. While with the Borinsky approach with tropical sector decomposition one might know beforehand those estimates, we do not rely solely on the Kaneko-Ueda sector decomposition strategy used there, thus we suggest another approach. First we use a smaller number of sampling points than the original {\tt maxeval} setting, for example, $100$ times less. After running such an integration we estimate the results in all sectors and then choose the {\tt maxeval} setting for each sector depending on the size of those results or error estimates.

We will provide benchmarks in section \ref{benchmarks}.

\textit{Complexity}: high;\\\indent
\textit{Parallelization}: close to linear for complicated integral up to the number of threads in use ({\tt NumberOfLinks} in {\tt Mathematica}), possibility to use even more cores on a supercomputer with the {\tt MPI} version, AVX processor instruction utilization, GPU usage;\\\indent
\textit{Related options}: a large number of options that are passed to the binaries {\tt CIntegratePool} or {\tt CIntegratePoolMPI};\\\indent
\textit{Possible issues}: with the increase of the {\tt maxeval} setting this part becomes the most lengthy stage of sector decomposition, but this might lead to a precision improvement.

\section{FIESTA installation and usage}
\label{installation}

\subsection{Installation}

{\tt FIESTA} is distributed via bitbucket. We no longer provide binary packages but do our best to provide an easy installation on different operating systems.

{\tt FIESTA} is known to work under Ubuntu 18.04 and 20.04 (also as part of the Windows subsystem for Linux, version 2), openSUSE Leap 15.2, macOS Big Sur 11.5, Centos 8. {\tt FIESTA} can be compiled with {\tt gcc} (minimal version is {\tt gcc-7}, known to work with {\tt gcc-11} and will be updated to match newer versions), {\tt icc} 2021 and {\tt clang-10}.

You will require some libraries to be installed to build FIESTA. We cannot provide instructions for every operating system but will use Ubuntu at least to name library package names and to simplify search. So on Ubuntu one will need

\begin{code} 
apt-get install git g++ cmake zlib1g-dev libmpfr-dev libgsl0-dev 
\end{code}

For some of the integration strategies and for the region approach to expansion also run

\begin{code} 
apt-get install qhull-bin
\end{code}

For the tensor train integrator also

\begin{code} 
apt-get install gfortran 
\end{code}

For the {\tt MPI} version also

\begin{code} 
apt-get install mpich
\end{code}

For the GPU version also

\begin{code} 
apt-get install nvidia-cuda-toolkit nvidia-cuda-dev
\end{code}

For documentation also

\begin{code} 
apt-get install doxygen
\end{code}

To get {\tt FIESTA} run

\begin{code} 
git clone https://bitbucket.org/feynmanIntegrals/fiesta.git
\end{code}

This command creates a {\tt fiesta} folder which can be renamed to something else without spoiling the installation. However please do not rename internal folders. Now move to the internal folder

\begin{code} 
cd fiesta/FIESTA5 
\end{code}

There run the {\tt configure script}

\begin{code} 
./congigure 
\end{code}

either with no options, or with some of the provided options (the script lists them anyway). There are the following possible options:

\begin{itemize}
 \item {\tt -$\-$-enable-qmc}: enable the quasi Monte-Carlo integrator;
 \item {\tt -$\-$-enable-tt}: enable the quasi Tensor Train integrator;
 \item {\tt -$\-$-enable-mimalloc}: switches memory allocator to mimalloc (faster);
 \item {\tt -$\-$-enable-avx}: turns on avx optimizations for evaluating in multiple points at a time by processor extensions; does not work on old architectures and on Mac OS X for lack of the aligned$\_$alloc function;
 \item {\tt -$\-$-math=NAME}: set up a proper {\tt Mathematica} binary, needed in case of multiple {\tt Mathematica} installations if a specific needs to be chosen;
 \item {\tt -cpp=NAME}: set the {\tt C++} compiler;
 \item {\tt -cc=NAME}: set the {\tt c} compiler needed for mathlink build;
 \item {\tt -mpicpp=NAME}: set the {\tt MPI C++} for the {\tt MPI} version.
 \end{itemize}

Then build the dependency libraries shipped with {\tt FIESTA} or downloaded automatically during the build (as usual, add -j for parallel build).

\begin{code} 
make dep 
\end{code}

Now build {\tt FIESTA}

\begin{code} 
make
\end{code}

In case you are building {\tt FIESTA} on a cluster with no {\tt Mathematica} installed for integration purposes you can skip the mathlink part of the build with

\begin{code} 
make nomath
\end{code}

Also on a cluster you might need the {\tt MPI} version.

\begin{code} 
make mpi
\end{code}

For support of graphical processing units for integration run

\begin{code} 
make gpu
\end{code}

After a successful build one can run some basic tests with

\begin{code} 
make test 
\end{code}

\noindent and some tests with {\tt Mathematica} with

\begin{code} 
make testmath
\end{code}

The code is covered with doxygen documentation. To build it run

\begin{code} 
make doc
\end{code}

\noindent and then open {\tt documentation/html/index.html} either directly in your browser or open it with a default browser with

\begin{code} 
make showdoc
\end{code}

\subsection{Basic usage}

The basic usage of {\tt FIESTA} is within {\tt Wolfram Mathematica}. One should load the file {\tt FIESTA5.m} for example with 

\begin{code}
 Get["FIESTA5.m"];
\end{code}

{\tt FIESTA5} is a {\tt Mathematica} package meaning it affects no variables from the global context. The public functions provided by this package are the following:

\begin{code}
UF[loop momenta, propagators, substitutions] 
\end{code}

\noindent generates the functions U and F and also the number of loops by the loop momenta, list of propagators and list of replacements. This function can be used in all following functions as a source of the first argument.

\begin{code}
SDEvaluate[\{U, F, loops\}, indices, order, options]
\end{code}

This function evaluates a Feynman integral; {\tt indices} is the set of indices, their number has to coincide with the number of propagators, {\tt order} is the requested expansion order by $\ep$ (or other variable in case of options). Here and below {\tt options} is a non-obligatory list of options passed to {\tt FIESTA}. All of them will be discussed in section~\ref{options}.

\begin{code}
SDEvaluateG[\{graph, external\}, \{U, F, loops\}, indices, order, options] 
\end{code}

\noindent is the same with SDEvaluate, but {\tt \{graph, external\}} is the most convenient syntax to provide a graph for the Speer strategy~\cite{Speer:1968:AR, Smirnov:2008aw};  here {\tt graph} is the number of pairs of vertices (numbers) and {\tt external} is the list of external vertices; the order of lines has to coincide with the order of indices. For details on the Speer strategy and examples see~\cite{Smirnov:2008aw}.

\begin{code}
SDExpand[\{U, F, loops\}, indices, order, expand$\_$degree, options] 
\end{code}

\noindent expands a Feynman integral; The syntax is the same with SDEvaluate, but there is also a parameter with the order of the expansion variable. Default expansion variable is t, but this can be changed by options. SDExpand can work only in case {\tt F} depends linearly on the expansion variable.

\begin{code}
SDExpandG[\{graph, external\}, \{U, F, loops\}, indices, order, expand$\_$degree, options] 
\end{code}

\noindent is a mixture of SDExpand and SDEvaluateG, it expands an integral with the help of Speer sectors.

\begin{code}
SDExpandAsy[\{U, F, loops\}, indices, order, expand$\_$degree, options]
\end{code}

\noindent expands the integral with expansion by regions~\cite{Beneke:1997zp, Smirnov:2002pj, Smirnov:2012gma}. The syntax completely coincides with the syntax of SDExpand, and it can work also with non-linear dependencies of {\tt F} on the expansion variable. The disadvantage is that an introduction of a regularization variable might be required. There is an example in the {\tt examples} folder ({\tt SDExpandAsyRegVar.m}) that was discussed in previous {\tt FIESTA} papers.

\begin{code}
SDEvaluateDirect[functions, degrees, order, deltas(optional), options] 
\end{code}

\noindent evaluates a non-loop integral, where degrees is the list of function powers; order is the expansion order; deltas, if provided, is the list of attached delta-functions. The example can be found in {\tt examples/SDEvaluateDirect.m}.

\begin{code}
SDExpandDirect[functions, degrees, order, expand$\_$degree, deltas(optional), options]
\end{code}

\noindent is similar for expansion. See {\tt examples/SDExpandDirect.m} for an example.

\begin{code}
SDAnalyze[{U, F, h}, degrees, dmin, dmax, options]
\end{code}

\noindent searches for values of space-time dimension resulting in possible ep-poles in the range from {\tt dmin} to {\tt dmax}.

{\tt SDIntegrate[options]} and {\tt GenerateAnswer[options]} are also public functions, but they will be explained in the advanced usage section.

Now {\tt FIESTA} has a large number of options, which can be set in two ways, either globally with something the help of {\tt SetOptions}:

\begin{code}
SetOptions[FIESTA, "NumberOfSubkernels" -> 4, "NumberOfLinks" -> 4];
\end{code}

\noindent or as the last parameter of any of the basic commands, for example, with

\begin{code}
SDEvaluate[\{x[1] + x[2] + x[3] + x[4], x[1] x[3] + x[2] x[4], 1\}, \{0, 1, 1, 1\}, 1, NumberOfSubkernels->4, NumberOfLinks -> 4];
\end{code}

In the latter case the option is applied only the the current command and does not influence other integrals that might be evaluated later. Please note that an option name in the global setting requires quotes and in the local does not need them.

{\tt FIESTA} is packaged with a number of examples. The basic ones are in the {\tt examples/examples.nb} file. There is also a number of {\tt *.m} files that one can test with {\tt math < examples/filename.m}.

\subsection{Advanced usage --- integrating separately}

The approach described above works in many cases, but if one is aiming for a good precision, then the default number of sampling points ($50000$) during the integration phase is in most cases not sufficient. Hence one needs to increase the number of sampling points. Moreover, one might not know in advance how many sampling points are going to be needed. Thus, not to rerun the algebraic preparation steps (everything but the integration) it is recommended to first run {\tt FIESTA} with the {\tt OnlyPrepare->True} option. This results in an integration database and the integration command, which is returned as a result. This command should now be run in a command line.

This approach is useful also in case the integration should be performed on a supercomputer with no {\tt Mathematica} installed. Then a database can be prepared elsewhere, transferred to the cluster and the integration should be performed there. Also one might wish to experiment with different integrators without rerunning the preparation part.

The key option use the possibility for separate integration is {\tt OnlyPrepare}. If {\tt DataPath} is not set, the database is located inside the {\tt temp} folder inside the {\tt FIESTA5} folder. The database name contains process id in the path so that multiple instances of {\tt FIESTA} do not interfere with each other. After the input database with integrals is produced, there are two ways to continue. One can either run the command

\begin{code}
SDIntegrate[options]
\end{code}

\noindent or, which is recommended, it is possible to run the integration from the command line without invoking {\tt Mathamtcica} using

\begin{code}
bin/CIntegratePool -$\-$-in temp/dbin.kch -$\-$-threads 4
\end{code}

Note: in general the command will differ and contain a process id in the database name as specified above; the number of threads is also provided as an example. One knows the database path and other options from the command returned by the {\tt Mathematica} part.

The binary {\tt bin/CIntegratePool} has a large number of options, and a big part of options set in {\tt Mathematica} are automatically translated to the command line options of {\tt bin/CIntegratePool}. For example, setting {\tt NumberOfLinks -> 4} is translated to {\tt -$\-$-threads 4}.

One can print out all options of {\tt bin/CIntegratePool} with the {\tt -$\-$-help} option, and all of them will be discussed in section~\ref{options}.

The two options having a maximum impact on the performance are the number of threads working in parallel {\tt -$\-$-threads} and the number of sampling points (being a common option for all integrators). The integration pool binary {\tt bin/CIntegratePool} is in fact not performing any integration, all that this binary does is launching a number (set by the threads option) of instances of {\tt bin/CIntegrateMP(?C)(?G)} (the binary name can either contain the {\tt C} and {\tt G} letters or not, see the explanation below), reads integrals from the input database and distributes them among integration processes, then collects results, prints them out and puts in the output database.

The result can either be picked from the command line output or in {\tt Mathematica} with the command

\begin{code}
GenerateAnswer[options]
\end{code}

As specified above, there are four integration binaries, either containing letters {\tt C} and {\tt G} in their names or not. Here {\tt C} stands for complex and {\tt G} stands for graphical processing unit --- GPU. The binaries with complex support also work in the case without complex numbers, but slower. The GPU binary is needed to take advantage of GPU if it exists.

The number of threads should normally be chosen equal to the number of cores on the computer --- the overhead for the pool binary is in general small, and {\tt FIESTA} efficiently loads the CPU. 

The number of sampling points is a common option for all integrators set with {\tt -$\-$-IntegratorOption maxeval:number} or with {\tt IntegratorOptions -> \{\{"maxeval", "number"\}\}} from {\tt Mathematica}, where {\tt number} is the required number of points ($50000$ by default). Normally the integration time is increased linearly with the number of sampling points after it becomes large enough to ignore the overhead for database communication and expression parsing. Increasing the number of sampling points is the most direct way to improve the quality of the result. Surely, there are multiple options affecting performance that will be discussed in section~\ref{options}.

When the number of sampling points becomes large, one might require a lot of time to run the integration. One of the ways to improve performance here is to take advantage of supercomputers and the {\tt MPI} version of {\tt FIESTA}. Switching to supercomputers is easy as it can be --- one needs to use the {\tt CIntegratePoolMPI} binary instead of {\tt CIntegratePool}. Of course, there might be a special command requires to utilize the {\tt MPI} infrastructure, for example,

\begin{code}
mpirun -n 128 bin/CIntegratePoolMPI -$\-$-in temp/dbin.kch
\end{code}

For details one should refer to the documentation of the supercomputer in use. In section~\ref{benchmarks} we will show how this can affect integration time.

\subsection{Special case --- individual integrals}

One might wish to analyze the integrals stored in the integration database or to integrate expressions obtained elsewhere. {\tt FIESTA} gives a possibility to get out integrals from the integration database with the {\tt bin/CIntegrateTool} utility. Start with the command

\begin{code}
bin/CIntegrateTool -$\-$-in temp/dbin.kch
\end{code}

It prints possible tasks --- individual Feynman integrals in the integration database. Normally there is a single task, but within the expansion by regions approach there might be multiple.

\begin{code}
bin/CIntegrateTool -$\-$-in temp/dbin.kch -$\-$-task 1 
\end{code}

This command prints the possible \textit{prefixes} for first Feynman integral, meaning possible sets of integrals with a common power of $\ep$ and power and logarithm power of the regularization variable. For example, the prefix for power $-2$ of $\ep$ and no regularization variable is $-2-\{0, 0\}$ and one can continue with

\begin{code}
bin/CIntegrateTool -$\-$-in temp/dbin.kch -$\-$-task 1 -$\-$-prefix "-2-\{0, 0\}"
\end{code}

\noindent or a shorter version in case of no regularization variable:

\begin{code}
bin/CIntegrateTool -$\-$-in temp/dbin.kch -$\-$-task 1 -$\-$-prefix -2
\end{code}

This command prints the number of sectors for this integral and prefix. One can not either print the number of functions in each sector with

\begin{code}
bin/CIntegrateTool -$\-$-in temp/dbin.kch -$\-$-task 1 -$\-$-prefix -2 --list
\end{code}

\noindent or the number of functions in an individual sector (for example, sector number $5$) with 

\begin{code}
bin/CIntegrateTool -$\-$-in temp/dbin.kch -$\-$-task 1 -$\-$-prefix -2 -$\-$-sector 5
\end{code}

The functions are individual integrals in a sector that appear due to the pole resolution stage and others.

Now with the number of functions one can either produce an integrand with one of those functions, for example, 

\begin{code}
bin/CIntegrateTool -$\-$-in temp/dbin.kch -$\-$-task 1 -$\-$-prefix -2 -$\-$-sector 5 -$\-$-function 1 
\end{code}

\noindent or make a combined expression with all of the functions in the given sector:

\begin{code}
bin/CIntegrateTool -$\-$-in temp/dbin.kch -$\-$-task 1 -$\-$-prefix -2 -$\-$-sector 5 -$\-$-all
\end{code}

The produced expression is in both cases exactly the input needed for the {\tt CIntegrateMP} binaries. It can be either sent directly there with pipe redirecting or saved to a file. Please note, that one should choose the complex or non-complex version of the binary. The binary without complex support will fail on complex numbers, the binary with complex support will work in the real case, but slower.
The following information in this section is related to the input of {\tt CIntegrateMP} binaries. The binary, when launches, waits for commands separated by line returns, the most important of those being {\tt Integrate}.

The structure of the integration expression after the {\tt Integrate} command is the following:

\begin{itemize}
 \item the number of integration variables --- integration dimension or number of positive indices minus one;
 \item the number of auxiliary functions;
 \item the auxiliary functions;
 \item the integrand itself.
\end{itemize}

All items should end with the semi-column (;) and an a new line. The whole expression ends with the vertical line symbol (|). The integration variables are noted as $x[i]$ with enumeration starting from $1$. The auxiliary functions are noted as $f[i]$ with enumeration starting from $1$. $p[a, b]$ stands for $a$ in power $b$, $l[a]$ stands for the natural logarithm of $a$, $P$ stands for $\pi$ and $G$ stands for EulerGamma, and also the PolyGamma[a,~b] notation is supported.

Some examples can be found in the {\tt examples} folder, all files with the {\tt int} extension.

The integration binary has no options but supports commands sent from standard input. Many integration options are passed from the integration pool and are translated to those options. For example, setting the {\tt maxeval} integrator parameter to $2000000$ is set as

\begin{code}
SetCurrentIntegratorParameter
maxeval
2000000
\end{code}

To see the full list of those options launch one of those binaries and run the {\tt Help} command. Since for normal usage this subsection is normally not needed and mostly is interesting for developers, we are not going to list the options here in the paper, but are ready to help on request.

\section{{\tt FIESTA} options}
\label{options}

Here we are going to list options of {\tt FIESTA}, both run from {\tt Mathematica} and when running integation pool from command line. The {\tt Mathematica} options are set with 

\begin{code}
SetOptions[FIESTA, "NumberOfSubkernels" -> 4, "NumberOfLinks" -> 4];
\end{code}

\noindent or as the last parameter of any of the basic commands, for example, with

\begin{code}
SDEvaluate[\{x[1] + x[2] + x[3] + x[4], x[1] x[3] + x[2] x[4], 1\}, \{0, 1, 1, 1\}, 1, NumberOfSubkernels->4, NumberOfLinks -> 4];
\end{code}

The {\tt CIntegratePool} options are set with, for example,

\begin{code}
bin/CIntegratePool -$\-$-threads 4
\end{code}

In this example the {\tt threads} options matches the {\tt NumberOfLinks}, and the {\tt NumberOfSubkernels} has no match in {\tt C++} since it is related only to the preparation of integrands performed purely in {\tt Mathematica}. On the other hand, some {\tt C++} options fine-tuning the integration process do not have their analogue in {\tt Mathematica}. The list of options with explanations follows (we tried to place most commonly used options first; the options that exist only in the {\tt C++} part of {\tt FIESTA} are in the end of the list):

\begin{itemize}
 \item {\tt NumberOfSubkernels} --- the number of {\tt Mathematica} subkernels used during all stages but integration in {\tt C++}; default value is $1$, it is reasonable to set it equal to the number of processor cores, however in complicated situations it should be made smaller to decrease RAM usage;
 \item {\tt NumberOfLinks} or {\tt -$\-$-threads} --- the number of working in parallel integration threads; default value is $1$, this option has no effect for {\tt bin/CIntegratePoolMPI};
 \item {\tt Strategy} --- sector decomposition strategy, possible recommended values are {\tt S2}, {\tt S}, {\tt KU}, {\tt KUS}, {\tt X}. Default value is {\tt S2}, default in previous versions was {\tt S}. Strategy {\tt KU} is expected to end with the smallest number of sectors, but might be too slow for a large number of positive indices. {\tt KUS} seems to be a good choice for a high number of positive indices balancing between speed and quality. {\tt X} is not guaranteed to terminate, but might be very effective;
 \item {\tt SectorSymmetries} --- whether sectors should be identified as symmetric, default value is {\tt True};
 \item {\tt ComplexMode} or {\tt -$\-$-complex} --- in case of {\tt True} turns on contour transformation needed in physical kinematics;  default version in {\tt Mathematica} is {\tt False}, the option is missing (it has no argument) in {\tt C++}; in {\tt C++} if set uses another binary with complex support;
 \item {\tt d0} --- the space-time dimension, default is {\tt 4};
 \item {\tt UsingC} --- whether to use {\tt C++} integration, default is {\tt True};
 \item {\tt RegVar} --- if set, uses an extra regularization variable for pole resolution and expansion, default value is {\tt None} meaning no regularization variable;
 \item {\tt ExpandVar} or {\tt -$\-$-expandVar} --- expansion variable name for expand modes, default value is {\tt t};
 \item {\tt XVar} --- coordinate variable name, default value is {\tt x};
 \item {\tt EpVar} --- small variable name, that is $(4-d)/2$, default value is {\tt ep};
 \item {\tt PMVar} --- error estimate variable name for output, default value is {\tt pm};
 \item {\tt Graph} --- internal option for Speer sector strategies, normally not set manually;
 \item {\tt NegativeTermsHandling} --- the setting for the way of preresolution. Default is {\tt Automatic}, meaning it is off in case {\tt ComplexMode -> True} and {\tt AdvancesSquares3} otherwise;
 \item {\tt PrimarySectorCoefficients} --- the coefficients at primary sectors, might be used to provide symetries manually or to split the integral into parts; default value is {\tt Automatic} meaning coefficient is equal to {\tt 1} at each sector. In case this option is set, {\tt FIESTA} does not search for symmetries between primary sectors. The value for the option should be a list of coefficients of length equal to the number of non-negative indices;
 \item {\tt OnlyPrepare} --- if set to {\tt True} makes {\tt FIESTA} only prepare databases and not run the integration step; default value is {\tt False};
 \item {\tt FixSectors} --- if set to {\tt True} runs the sector decomposition stage with no parallelization making the numbering of sectors coincide for different runs; useful for debugging; default value is {\tt False};
 \item {\tt MixSectors} --- if set to a positive number $n$, combines results of different $n$ sectors into one; might be needed in rare cases; default value is {\tt 0};
 \item {\tt SectorSplitting} --- if set to {\tt True} makes {\tt FIESTA} search for possible zeros of $F$ due to $x[i]->1$ and bisect sectors by such variables; default value is {\tt False}; should be tried as {\tt True} in case of convergence problems in complex mode;
 \item {\tt MinimizeContourTransformation} --- if set to {\tt True} (default value) makes contour transformation only for those variables that we believe to be responsible for sign change; might be set to {\tt False} in case of convergence problems in complex mode;
 \item {\tt ContourShiftShape} --- after balancing of shifts we allow maximum lambda equal to {\tt ContourShiftShape}; larger values let the imaginary shift be larger than the original real value; default value is {\tt 1};
 \item {\tt ContourShiftCoefficient} --- after the best lambda is found, it is multiplied by this coefficient; default value is {\tt 1};
 \item {\tt ContourShiftIgnoreFail} --- in case the contour transformation fails for some reasons to find a reasonable lambda, and this option is at its default value, {\tt False}, {\tt FIESTA} reports an error and stops working; if one wishes to continue without contour transformation in this case, this option should be set to {\tt True};
 \item {\tt FixedContourShift} --- if this option is set to {\tt True}, then after balancing the following steps are not performed and this value is taken for lambda;
 \item {\tt LambdaIterations} --- the number of numerical checks performed at each of the substeps during the contour transformation, default value is {\tt 1000}; greater values slow the stage down, but might result in a better lambda search improving integration convergence;
 \item {\tt LambdaSplit} --- the number of lambda values checked at the substep searching for best possible lambda after finding the maximum; default value is {\tt 4};
 \item {\tt ChunkSize} --- the number of functions passed in a chunk to subkernels on all stages in {\tt Mathematica}, default value if {\tt 1}, can be increased in case of simple integrals and poor performance (CPU not loaded well), however making it bigger increases RAM usage, and the value should be still much smaller than the number of sectors, otherwise performance will go down;
 \item {\tt OptimizeIntegrationStrings} --- if set to {\tt True} runs {\tt Mathematica} expression optimization at expression generation stage as it was done in old versions of {\tt SecDec}; default value is {\tt False} and this stage might be too slow with the {\tt True} setting;
 \item {\tt AnalyzeWorstPower} --- if set to {\tt True} analyzes expressions saves to strings trying to recursively find worst behaviour of the $x[i]^{-a}$ type. After that the expression is ``multiplied'' by such a monomial and inverse one not changing the expression but letting the {\tt C++} part know such a behavior. The default setting was {\tt True} for previous expressions (in face there was no option), now in is {\tt False}. The ``worst monomial'' knowledge is needed to properly decide which point should use multi-precision calculations, however the direct approach of power analysis inside the {\tt C++} seems to work; currently we know no single example where this option should be set to {\tt True} but this could be a test option in case of completely incorrect results with even larger error estimates; please notify us in case this option helps to improve results;
 \item {\tt ZeroCheckCount} --- if set to a non-zero value makes {\tt Mathematica} at expression generation stage perform that many checks substituting different numerical values into each function and checking if the result is equal to zero; small values might result in incorrectly deciding that a function is equal to zero totally, large values lead to low performance at this stage; we believe that currently there is no need to change the default {\tt 0} value but in case the functions in your sector often have hidden zeros this could be the case;
 \item {\tt ExpandResult} --- whether the final result after being collected from result of different orders should be expanded in $\ep$ and other small variables; default value is {\tt True};
 \item {\tt FIESTAPath} --- path to the folder containing {\tt FIESTA5.m} and other files, normally is detected automatically when {\tt Mathematica} loads {\tt FIESTA5.m} and should not be changed;
 \item {\tt DataPath} --- option setting the path for the databases with integrals; this option is translated to the {\tt -$\-$-in} option (see below) when integration pool is called; possible values: a specified path, {\tt Automatic} meaning a process id depending value inside the local {\tt temp} folder inside the {\tt FIESTA} folder and {\tt Default} meaning {\tt temp/dbin.kch}; default value is {\tt Automatic}; in case one wishes the database files to be cleaned up in any case including canceling a job he might consider changing directory to a temporary folder and providing a local database path before starting {\tt FIESTA};
 \item {\tt BucketSize} or {\tt -$\-$-bucket} --- specify kyotocabinet bucket value, might be important for speed for databases with a large number of entries, default value is {\tt 27}; 
 \item {\tt NoDatabaseLock} --- if set to {\tt True} makes {\tt Mathematica} not lock database files on disk; this might be needed in rare cases due to problems with network disks; default value is {\tt False};
 \item {\tt RemoveDatabases} --- whether to remove databases after integration is over; default value is {\tt True};
 \item {\tt SeparateTerms} or {\tt -$\-$-separateTerms} --- each sector might contain several expressions due to singularity resolution, this option makes terms inside a sector to be sent for integration separately, not mixed; the option might have a great impact on performance (!), however it cannot be predicted in advance, whether in should be set or not; default version in {\tt Mathematica} is {\tt False}, the option is missing (it has no argument) in {\tt C++};
 \item {\tt BalanceSamplingPoints} or {\tt -$\-$-balanceSamplingPoints} --- if set to {\tt True} in {\tt Mathematica} (or just set in {\tt C++}) first runs an estimation run with a small number of sampling points, only then the final run; default version in {\tt Mathematica} is {\tt False}, the option is missing (it has no argument) in {\tt C++};
 \item {\tt BalanceMode} or {\tt -$\-$-balanceMode} --- sets the mode used for balancing, one of {\tt realValue}, {\tt realError}, {\tt imaginaryValue}, {\tt imaginaryError}, {\tt normValue} and {\tt normError}, comparing for each of the integrand results either results or error estimates of real parts, imaginary parts of norms; default value in {\tt Mathematica} is {\tt Automatic} meaning {\tt normError};
 \item {\tt BalancePower} or {\tt -$\-$-balancePower} --- the values from {\tt BalanceMode} are taken into this power, measured and compared with each other to choose the number of sampling points for the final run; default value in {\tt Mathematica} is {\tt Automatic} meaning {\tt 0.5};
 \item {\tt ResolutionMode} --- one of resolution modes at the pole resolution stage, default value is {\tt Taylor}, other possible values are {\tt IBP0} and {\tt IBP1};
 \item {\tt AnalyticIntegration} --- option valid only for {\tt SDExpandAsy}, if set to {\tt True} tries to analytically get rid of some integration variables after regions detection, {\tt False} by default;
 \item {\tt OnlyPrepareRegions}  --- option valid only for {\tt SDExpandAsy}, if set to {\tt True}, only reveals regions and prints them, {\tt False} by default;
 \item {\tt AsyLP} --- option valid only for {\tt SDExpandAsy}, now {\tt True} by default, using Lee-Pomeransky representation for regions search, which is much faster;
 \item {\tt PolesMultiplicity} --- option having sense only for {\tt SDAnalyze}, if set to {\tt True} lists poles with those multiplicities; default value is {\tt False};
 \item {\tt ExactIntegrationOrder} --- option having sense only in case of integration in {\tt Mathematica} ({\tt UsingC -> False}), if set to number tries to integrate analytically up to provided order; default value is {\tt -Infinity} meaning no analytic integration attempts;
 \item {\tt ExactIntegrationTimeout} --- option having sense only in case of integration in {\tt Mathematica} ({\tt UsingC -> False}), sets the time limit for an analytic integration attempt for a function in seconds; default value is {\tt 10};
 \item {\tt GPUIntegration} or {\tt -$\-$-gpu} --- makes the pool use the GPU binary for all (or some) threads;  default value in {\tt Mathematica} is {\tt False}, by default the option is not set (it has no argument) in {\tt C++};
 \item {\tt NoAVX} or {\tt -$\-$-NoAVX} --- if set to {\tt True} in {\tt Mathematica} switches off AVX optimization for function evaluation, has sense only if configured with {\tt -$\-$-enable-avx}; default value in {\tt Mathematica} is {\tt False}, by default the option is not set (it has no argument) in {\tt C++};
 \item {\tt Precision} or {\tt -$\-$-precision} --- precision in digits used in results; default value is {\tt 6};
 \item {\tt ReturnErrorWithBrackets} --- if set to {\tt True} the error estimate variable is printed as {\tt pm[i]} instead of {\tt pmi} for integer {\tt i}, default value is {\tt False},
 \item {\tt Integrator} or {\tt -$\-$-Integrator} --- sets the integrator provided by one of the integration libraries, default value is {\tt vegasCuba};
 \item {\tt IntegratorOptions} or multiple instances of {\tt -$\-$-IntegratorOption} --- sets options for the chosen integrator; in {\tt Mathematica} the value should be a list of pairs (option name and value); in {\tt C++} the syntax is {\tt -$\-$-IntegratorOption <option$\_$name>:<option$\_$value>} for each parameter; default value in {\tt Mathematica} is {\tt Automatic} meaning no change for default integrator options; their list with values is printed when {\tt FIESTA} starts working;
 \item {\tt CIntegratePath} or {\tt -$\-$-cIntegratePath} --- provides a path to the CIntegrate binary. Overrides {\tt GPUIntegration/-$\-$-gpu} and {\tt ComplexMode/-$\-$-complex} options used for detection of proper path; default value in {\tt Mathematica} is {\tt Automatic} meaning no path passed to the integration pool;
 \item {\tt MPSmallX} or {\tt -$\-$-MPSmallX} --- sets the coordinate value (all x) for the small point where the worst monomial is measured; default value in {\tt Mathematica} is Automatic, meaning $0.001$;
 \item {\tt MPThreshold} or {\tt -$\-$-MPThreshold} --- defines the limit for the worst monomial in MPSmallX test point, so that for smaller products MPFR is turned on; default value in {\tt Mathematica} is Automatic, meaning $1E-9$;
 \item {\tt MPMin} or {\tt -$\-$-MPMin} --- sets the limit for the worst monomial on where to stop using default precision and switch to shifts; default value in {\tt Mathematica} is Automatic, meaning $1E-48$;
 \item {\tt MPPrecisionShift} or {\tt -$\-$-MPPrecisionShift} --- sets additional bits for mpfr precision when performing calculations in a point; default value in {\tt Mathematica} is Automatic, meaning $38$;
 \item {\tt MathematicaBinary} or {\tt -$\-$-Math} --- provides path to the math binary for the integrators, makes evaluation of non-predefined constants possible; default value in {\tt Mathematica} is {\tt None} meaning no value; if set to {\tt Automatic} makes {\tt Mathematica} detect the value from the build configuration ({\tt -$\-$-math=...}) and pass it to {\tt C++}; Note: with personal edition of {\tt Mathematica} one cannot launch multiple instances of mathkernel at a time. So one should run examples requiring calls for polygamma evaluation from {\tt C++} first with {\tt OnlyPrepare -> True} and then with $1$ thread;
 \item {\tt QHullPath} --- the path to call the qhull binary needed for the {\tt KU} strategy and {\tt ASY} modes; default value is {\tt qhull};
 \item {\tt DebugParallel} --- if set to {\tt True}, prints information on the load of subkernels, default value is {\tt False};
 \item {\tt DebugMemory} --- if set to {\tt True}, prints information on memory usage, default value is {\tt False};
 \item {\tt DebugAllEntries} --- if set to {\tt True}, prints all intermediate expressions (please be careful setting it on especially in a {\tt Mathematica} notebook since it might produce huge amounts of output), default value is {\tt False};
 \item {\tt DebugSector} --- if set to a positive number, restricts {\tt FIESTA} to only working in the specified sector; should be set together with {\tt FixSectors -> True}; default value is {\tt 0};
 \item {\tt -$\-$-in} --- specifies input database with integrals; if name does not end with .kch, the extension is appended automatically, if name ends with in.kch, automatically sets the output database with the same value but ending with out.kch in case it was not set; should be always set for the integration pool;
 \item {\tt -$\-$-out} --- specifies output database. If name does not end with .kch, the extension is appended automatically; obligatory if the name of the input database does not end with {\tt in} or {\tt in.kch};
 \item {\tt -$\-$-debug} --- prints results for all sectors and functions there; this option is not related to different debug options for {\tt Mathematica};
 \item {\tt -$\-$-gpuThreadsPerNode} --- only if {\tt -$\-$-gpu} set, makes only <value> instances per node use GPU;
 \item {\tt -$\-$-gpuPerNode} --- only if {\tt -$\-$-gpu} set, makes instances using GPU use different GPUs (by default all use GPU number 0);
 \item {\tt -$\-$-fromMathematica} --- makes pool write intermediate results in special files for communication with {\tt Mathematica} instead of printing results and saving them to output database;
 \item {\tt -$\-$-queueSize} --- sets the queue size of jobs of integrands and results; large sizes might be needed in case of many threads or MPI so that the queue does not suddenly get empty; default value is {\tt 32}; 
 \item {\tt -$\-$-printIntegrationCommand} --- prints the command the pool was called with to stdout before working;
 \item {\tt -$\-$-Preparse} --- runs the integrators first with a parse check only then with real integration, might be useful for faster debugging of parse problems;
 \item {\tt -$\-$-task} --- sets the task number of integrals to be integrated from the input database; for all modes but expansion with regions there is only one task; see instructions for {\tt CIntegrateTool} to get the list of tasks;
 \item {\tt -$\-$-prefix} --- sets the prefix for integrals to be integrated from the input database; prefixes are integration orders of form {\tt 0-\{0, 0\}}; see instructions for {\tt CIntegrateTool} to get the list of prefixes;
 \item {\tt -$\-$-onlyPrepare} --- with this option integration pool only creates the output database and populates it with dummy result entries. This option might be useful to create a database first and fill in with separate task/prefix/part jobs later. Note: this option has nothing common with the {\tt OnlyPrepare} option in {\tt Mathematica};
 \item {\tt -$\-$-part} --- with <value>=<part number>/<parts number> run this option after preparing the out database consequently for <part number> = 1 .. <parts number> to get the final result in the out database;
 \item {\tt -$\-$-continue} --- makes it possible to continue interrupted evaluation; stores all intermediate results in out database;
 \item {\tt -$\-$-continueKeep} --- in adition keeps intermediate results in output database after calculation is over;
 \item {\tt -$\-$-noOutputDatabase"} --- prevents pool from saving results to an output database, only keeps printing them;
 \item {\tt -$\-$-printStatistics} --- print average number of sampling points used for each prefix;
 \item {\tt -$\-$-Test} --- only runs a sample integration test, no real integration is perfomed;
 \item {\tt -$\-$-FTest} --- instead of the integration picks the F-function from the database and puts integrators in testF mode. In this mode the functions are also integrated (however the result of F-integration is meaningless). However the main purpose is a test that F functions always have same sign (needed in complex mode). In case they are not, an x value is printed and an error is produced. This mode cn be used to check whether the contour transformation was OK if you observe convergence problems in complex mode;
 \item {\tt -$\-$-Native} --- forces integrators to use native evaluations only, no MPFR; faster, but might lead to errors;
 \item {\tt -$\-$-MPFR} --- forces integrators to use MPFR evaluations only, no native arithmetic; much slower;
 \item {\tt -$\-$-NoOptimization} --- switches off triad optimization in CIntegrate, only for debugging reasons.
\end{itemize}

\section{Examples and benchmarks}
\label{benchmarks}

In this section we are going to provide some benchmarks showing performance of {\tt FIESTA} compared to old versions and also demonstrating differences of performance between different options.

\subsection{Example without complex numbers, comparing versions, strategies and MPI}

\label{F1}
Let us first consider master integrals for the family of Feynman integrals corresponding to the graph in equation (6) from~\cite{Lee:2021uqq}. Two external momenta are on the light cone ($p_1^2=p_2^2=0$) and all the masses are zero. The analytical result for the integral with all the first 12 indices equal to one and all the other indices equal to zero is given by the right-hand side of this equation. Analytical results
for all the other master integrals will be presented in a future publication.

Those integrals do not require contour transformation. However integrals with a larger number of positive indices have poles of multiple orders that require pole resolution, and as a result the preparation might take much time, and the numerical integrals are quite complex. We produce results up to $\ep^0$.

We checked master integrals with new and old versions of {\tt FIESTA}. The tests were performed on a laptop with a 4-core AMD Ryzen 7 3750H processor and are provided in table~\ref{versions_strategies}. The time is measured in seconds, the table header contains integral numbers and corresponding number of positive indices. 
We consider four typical integrals with the number of positive indices from 7 to 10. The numbers of integrals do not make much sense here because we do not present the full list of the master integrals but the definitions of these typical integrals can be directly 
obtained from the Mathematica commands in the file with examples distributed together with the code.
We used the version 12 of {\tt Wolfram Mathematica} (note: {\tt Mathematica 12.0} has a bug resulting in rare crashes of {\tt FIESTA}, version {\tt 12.3} is recommended).

\begin{table}[ht]
\begin{center}
\begin{tabular}{ |c|c|c|c|c| } 
 \hline
 integral                       & 9 (7) & 19 (8) & 42 (9) & 58 (10)\\ 
 \hline
 FIESTA4 - strategy S           &     1 &     8  &     37 &    865 \\ 
 FIESTA4 - prepare after S      &    20 &    74  &    337 &   3985 \\ 
 FIESTA5 - strategy S            &     1 &     6  &     35 &    884 \\ 
 FIESTA5 - prepare after S      &     9 &    26  &     95 &   2528 \\ 
 FIESTA5 - strategy S2 (default) &     1 &     5  &     29 &    405 \\ 
 FIESTA5 - prepare after S2     &     9 &    26  &    101 &   2570 \\ 
 FIESTA5 - strategy KUS          &     8 &    17  &     58 &    276 \\ 
 FIESTA5 - prepare after KUS    &     8 &    25  &     81 &   1250 \\
 FIESTA5 - strategy KU           &     9 &    19  &    265 &      * \\ 
 FIESTA5 - prepare after KU     &     7 &    19  &     58 &      - \\ 
 \hline
\end{tabular}
\end{center}
\caption{Comparing versions and strategies. The labels in column heads designate integral numbers and numbers of positive indices. Time is measures in seconds.\label{versions_strategies}}
\end{table}

The result could not be achieved for integral 58 with strategy KU taking forever. 

We also measured the number of sectors obtained for different strategies. The number of sectors generally defines the complexity of stages following sector decomposition, the less sectors one has, the simpler is the task in general. Here are the numbers in pairs --- all sectors and unique sectors for different strategies:

\begin{itemize}
 \item integral  9: 574/461  (KU), 650/521 (KUS), 652/407 (S2), 652/407 (S);
 \item integral 19: 1240/785  (KU), 1589/968 (KUS), 1676/865 (S2), 1726/827 (S);
 \item integral 42: 2733/1612 (KU), 3586/2039 (KUS), 4437/2569 (S2), 4025\\/2144 (S);
 \item integral 58: no result (KU), 15982/15982 (KUS), 27542/27542 (S2), 38596/38596 (S).
\end{itemize}

There is no clear winner by the number of sectors between S2 and S, but for higher number of positive indices S2 wins. Moreover, S2 is much faster in this case. KUS produces much less sectors and is faster with more positive indices but slow when the number of positive indices is smaller. KU produces the minimal number of sectors but is too slow and fails with $10$ or more indices. And anyway the new {\tt FIESTA} is faster than the old one, especially with a proper choice of strategies.

The example files (including diagram and master integral definitions) are shipped with {\tt FIESTA}. To prepare an integration database for this example, run

\begin{code}
 examples/F1/generate$\_$db.sh NUMBER CORES STRATEGY
\end{code}

\noindent where NUMBER is the number of master integral (all master integrals are listed in {\tt examples/F1/F1-masters.m}), CORES is the number of cores (translated to the {\tt NumberOfSubkernels} option) and STRATEGY is the chosen sector decomposition strategy. For example,

\begin{code}
 examples/F1/generate$\_$db.sh 58 4 KUS
\end{code}

The CORES and the STRATEGY options can be not filled, default is $1$ and {\tt S2}. 

Now let us provide the integration comparison. After generating integrand databases one can run the integration with the following command:

\begin{code}
 examples/F1/integrate$\_$db.sh NUMBER CORES MAXEVAL
\end{code}

\noindent where NUMBER is the number of master integral, CORES is the number of cores (translated to the {\tt NumberOfSubkernels} option) and MAXEVAL is the maximum number of sampling points. For example,

\begin{code}
 examples/F1/integrate$\_$db.sh 58 4 500000
\end{code}

In our tests we used $500000$ sampling points instead of the default $50000$. Increasing it further leads to about linear growth of time and provides no extra information for tests.

We chose the KUS strategy for the new version of {\tt FIESTA} based on the previous table and compared integration time with the old version (the results and error estimates are similar). We got the following timings (in seconds):

\begin{itemize}
 \item integral 9: {\tt FIESTA4} --- 21, {\tt FIESTA5} --- 13;
 \item integral 19: {\tt FIESTA4} --- 88, {\tt FIESTA5} --- 32;
 \item integral 42: {\tt FIESTA4} --- 245, {\tt FIESTA5} --- 73;
 \item integral 58: {\tt FIESTA4} --- 5333, {\tt FIESTA5} --- 2043;
\end{itemize}

This shows a considerable speedup of the new version even in the case without physical kinematics (complex numbers). One might notice a smaller speedup for integral $58$ compared to $42$, this is due lack of symmetries for this integral between sectors. The answers are known analytically and can be checked with

\begin{code}
 examples/F1/check$\_$result.sh NUMBER
\end{code}

We also used this example to test the MPI mode of {\tt FIESTA}. We used integral $58$ and the {\tt separateTerms} mode --- this was essential for a proper parallelization with a large number of workers, otherwise some integrals were dominating time usage. The test was performed on the cluster of the Institute for Theoretical Particle Physics at the Karlsruhe Institute of Technology, and the command was

\begin{code}
 srun -n $\$$n ./bin/CIntegratePoolMPI \\
      -$\-$-in ./examples/F1/temp/db$\_$58$\_$in -$\-$-separateTerms \\
      -$\-$-Integrator vegasCuba -$\-$-IntegratorOption maxeval:10000000 \\
      -$\-$-IntegratorOption epsrel:0.01 
\end{code}

Note: the command to use the MPI system on a cluster might differ on different clusters, please refer to the local documentation.

We obtained the timings with different number of cores and machines in use, which can be seen in table~\ref{mpi}.

\begin{table}[ht]
\begin{center}
 \begin{tabular}{|c|c|c||c|c|c|}
 \hline
 cores & machines & time & cores & machines & time \\
 \hline
2 & 1 & 4173 & 55 & 4 & 165 \\
3 & 1 & 2451 & 79 & 2 & 100 \\
4 & 1 & 2071 & 79 & 3 & 155 \\
6 & 1 & 1268 & 79 & 4 & 130 \\
9 & 1 & 845  & 114 & 3 & 74 \\
13 & 1 & 554 & 114 & 5 & 99 \\
18 & 1 & 424 & 114 & 6 & 101 \\
18 & 2 & 421 & 165 & 3 & 60 \\
27 & 1 & 256 & 165 & 7 & 100 \\
27 & 2 & 316 & 165 & 8 & 82 \\
27 & 3 & 310 & 237 & 8 & 67 \\
38 & 1 & 201 & 237 & 5 & 47 \\
38 & 2 & 247 & 342 & 6 & 44 \\
\hline
\end{tabular}
 \caption{MPI example. The time is measured in seconds. The timings are unstable due to the heterogeneous nature of the cluster in use\label{mpi}}
 \end{center}
\end{table}

\subsection{Physical kinematics, comparing versions and options of the new {\tt FIESTA}}

We used FIESTA to check analytic results for the master integrals for the second type of planar contributions to the massive two-loop Bhabha scattering in quantum electrodynamics.

These analytic results were obtained in a recent paper~\cite{Duhr:2021fhk} using differential equations with canonical bases.

This is a two-loop example, but it already demonstrates well the advancement from {\tt FIESTA4} to {\tt FIESTA5}. Here there is no difference between sector decomposition strategies. We produce results up different orders depending on the integral (as required for the physical check). We consider four typical integrals with the number of positive indices from 4 to 7. The definitions of these integrals also can directly be obtained from the Mathematica commands in the file with examples distributed together with the code.

\begin{table}[ht]
\begin{center}
 \begin{tabular}{|c|c|c|c|c|}
 \hline
 integral                  & 10 (4) &  23 (5) &  37 (6)   & 40 (7)\\
 \hline
 requested order           & 2      &  1      &  0        & 0   \\
 FIESTA4 - preparation     & 8      &  34     &  273      & 1242 \\
 FIESTA4 - integration     & 314    &  628    &  687      & 3269 \\
 FIESTA5 - preparation     & 2      &  3      &  9        & 10    \\ 
 FIESTA5 - integration     & 53     &  82     &  77       & 177    \\
\hline
 \end{tabular}
 \caption{Integration comparing in physical kinematics. The time is measured in seconds.\label{pl2}}
 \end{center}
\end{table}

The results can be seen in table~\ref{pl2}. Here we took $1000$ times more sampling points compared to the default value --- $50$ million sampling points. We also used the default {\tt vegasCuba} integrator. First of all, it is clearly obvious that {\tt FIESTA5} completely outperforms {\tt FIESTA4} even on this two-loop example. The integration results are known by other means and the results of {\tt FIESTA5} are much more close to them than the results of {\tt FIESTA4}.

However, timings are not the only problem. For example, for the final part of integral $40$ {\tt FIESTA5} with these options returns $13.1300 - 18.3537*I$ while the known result is $13.0927 - 18.3447*I$ being quite close, while {\tt FIESTA4} with the same number of sampling points produces $7.86407  + -1.79244 * I$ which is completely far from the answer. A better answer can be achieved with the increase of the number of sampling points, but this might take too much time.

These benchmarks with {\tt FIESTA5} can also be reproduced with

\begin{code}
 examples/pl2/generate$\_$db.sh NUMBER CORES STRATEGY
\end{code}

\noindent for database generation, with

\begin{code}
 examples/pl2/integrate$\_$db.sh NUMBER CORES MAXEVAL
\end{code}

\noindent for integration and with 

\begin{code}
 examples/pl2/check$\_$result.sh NUMBER
\end{code}

\noindent to check results.

Now let us also demonstrate the balancing of the number of sampling points of different integrals. The basic integration for integral $40$ is performed with

\begin{code}
 bin/CIntegratePool -$\-$-in ./examples/pl2/temp/db$\_$40$\_$in -$\-$-threads 4 -v -$\-$-IntegratorOption maxeval:50000000 -$\-$-complex -$\-$-separateTerms 
\end{code}

In takes 177 seconds and results in an error equal to $0.0372 - 0.0090 I$ for the final part. However we can take advantage of balancing and run it with

\begin{code}
 bin/CIntegratePool -$\-$-in ./examples/pl2/temp/db$\_$40$\_$in -$\-$-threads 4 -v -$\-$-IntegratorOption maxeval:50000000 -$\-$-complex -$\-$-separateTerms -$\-$-balanceSamplingPoints
\end{code}

Then it takes 66 seconds, and the error is only slightly larger being equal to $0.0408 - 0.01480 I$.

On the other hand, we can use this example to demonstrate, how important the improvements with AVX instructions are by running the following command:

\begin{code}
 bin/CIntegratePool -$\-$-in ./examples/pl2/temp/db$\_$40$\_$in -$\-$-threads 4 -v -$\-$-IntegratorOption maxeval:50000000 -$\-$-complex -$\-$-NoAVX 
\end{code}

It takes 342 seconds (with the same result as with AVX of course) being almost twice longer.

\subsection{Contour decomposition options and integrators}

Let us consider one more example having two linear propagators with dimension of momentum-space close to $3$, not $4$. This is an example with two heavy quark propagators and three massless propagators, which appears as a master integral in the computation of the two-loop static potential in three-dimensions (this integral first appeared in the context of~\cite{Kalin:2020fhe}).

If one simply tries to prepare the database with

\begin{code}
 SDEvaluate[UF[\{l1, l2\}, \{-l1 u, l2 u, -(l1 + l2 - q)$^2$, -(l1 - q)$^2$, -(l2 - q)$^2$\}, \{q$^2$ -> -1,  u$^2$ -> -1, u*q -> 0\}], \{1, 1, 1, 1, 1\}, 0, d0 -> 3, ComplexMode->True]
\end{code}

\noindent then the following integration produces unstable results --- the answer for the finite part does not seem to converge with the growth of the number of sampling points. There are a few things one should try in this case, and one of them is to add

\begin{code}
 SectorSplitting -> True
\end{code}

\noindent to the list of options. This option leads to search for singularities at $1$, not also at $0$ (see the corresponding section for details. As a result we can see the following lines in the log file:

\begin{code}
Counting different sectors: 32 terms.
Searching for x[i]->1 singularities..........58 terms.
Additional sector decomposition..........72 terms.
\end{code}

However, this still does not help, the result still does not converge. Now the idea is to check whether perhaps the contour transformation went out of bounds resulting in a positive imaginary part of $F$. This can be checked with the {\tt FTest} option:

\begin{code}
bin/CIntegratePool -$\-$-in ./examples/lin2/temp/db$\_$in -$\-$-threads 4 --Integrator vegasCuba -$\-$-IntegratorOption maxeval:500000 -$\-$-complex -$\-$-separateTerms -$\-$-FTest
\end{code}

This function performs the integration of $F$ after transformation, which has no particular meaning, but also watches for the sign. And as a result one can see the following messages in the log file (many of those):

\begin{code}
Integrating.......\{0.647407, 0.996573, 0.983618, 0.997353, \} result: \{1.248828898651168,0.026095569042321\}
\{0.792542, 0.985277, 0.992735, 0.989432, \} result: \{1.613810831804675,0.040549561921149\}\\
\{0.658998, 0.996897, 0.981823, 0.996996, \} result: \{1.157437564597312,0.443214740778332\}\\
\{0.679699, 0.977365, 0.918743, 0.997780, \} result: \{0.826571082996057,0.150831603598730\}\\
\{0.842319, 0.993608, 0.858646, 0.998647, \} result: \{0.996115222194668,0.023579557535055\}\\
\end{code}

This confirms that the reason is in a contour transformation that is too large and can be fixed by making the {\tt ContourShiftCoefficient} option smaller, for example, equal to $1/2$. This finally helps, and the setting can be seen in {\tt examples/lin2/generate$\_$db.m}. The database can be generated with

\begin{code}
 examples/lin2/generate$\_$db.sh CORES STRATEGY
\end{code}

\noindent integrated with

\begin{code}
 examples/lin2/integrate$\_$db.sh CORES MAXEVAL INTEGRATOR
\end{code}

\noindent and checked with

\begin{code}
 examples/lin2/check$\_$result.sh
\end{code}

However, sometimes the results are technically correct, but the convergence is too slow. This could be the case that the contour transformation is too small, and here the {\tt ContourShiftCoefficient} option should be increased.

We also use this example to compare different integrators. We checked this example from $50$ thousand sampling points to $5$ billion sampling points with two integrators, {\tt vegasCuba} and {\tt quasiMonteCarlo}. To force the integrators produce more digits we set a small {\tt epsrel} setting for both integrators and also used the balancing mode of {\tt FIESTA}. We do not provide the comparing of balancing and no balancing here, but one can easily check that balancing is significantly faster, especially for a large number of sampling points. The results are shown at table~\ref{integrators}.

\begin{table}[ht]
\begin{center}
 \begin{tabular}{|c|c|c|c|c|}
 \hline
 Sampling points   & vegas - time &  vegas - error & qmc - time & qmc - error   \\
 \hline
       50 000      &          1   &  1.31565614    &         28 &  0.02736893   \\
      500 000      &          2   &  0.15482110    &         31 &  0.01538130   \\
    5 000 000      &         16   &  0.01413316    &        110 &  0.00272567   \\
   50 000 000      &        148   &  0.00188693    &        868 &  0.00005208   \\
  500 000 000      &       1420   &  0.00045132    &       3386 &  0.00000185   \\
5 000 000 000      &      14649   &  0.00009603    &       9246 &  0.00000044   \\
 \hline
 \end{tabular}
 
 \caption{Integrator comparing\label{integrators}}
 \end{center}
\end{table}

The results show that the quasi monte carlo method produces better results when a large number of sampling points is used. It first starts slower due to the overhead, the larger number of sampling points it takes (exceeding maxeval) and also due to a larger portion of sampling points that need to be evaluated with the use of MPFR. However it produces better results and in the end it becomes faster for the reason that it can stop with most of the sector integrals early enough concentrating only on important one.

In practice one should choose the integrator wisely depending on parameters and the accuracy goal that should be achieved. For example, in subsection~\ref{F1} switching to QMC would only make the integration take more time.

\section*{How to cite}

{\tt FIESTA5} depends on a number of different libraries. Apart from citing this paper this would make sense to cite also:
\begin{itemize}
 \item the Cuba library~\cite{Hahn200578} as the default integrator;
 \item the QMC integrator~\cite{BOROWKA2019120} in case it is used;
 \item the tensor train integrator~\cite{Vysotsky2021} in case it is used.
\end{itemize}

\section*{Conclusion}

We presented a new release of {\tt FIESTA}. The benchmarks show that it is much more powerful than the old version of {\tt FIESTA}, outperforming it from $2-3$ times to $100$ times or even more working where the previous version was not capable to provide a result. This is achieved by new sector decomposition strategies, improved contour transformation, new integrators and multiple integration optimizations. We have more plans on how the code of {\tt FIESTA} should be developed to improve performance.

\section*{Acknowledgements}

The work is supported by Russian Ministry of Science and Higher Education, agreement No. 075-15-2019-1621. We would like to thank V.~Smirnov, M.~Steinhauser and V.~Shtabovenko for long-lasting beta-testing of the new {\tt FIESTA}, V.~Magerya for help with building and setting up the {\tt MPI} version of {\tt FIESTA} and testing it at the KIT cluster, G.~Mishima for help with building {\tt FIESTA} on Max OS X,  V.~Smirnov, M.~Steinhauser and Y.~Schroeder for a careful reading of the draft of this paper.

\bibliographystyle{elsarticle-num-names}
\bibliography{FIESTA5, asmirnov}
\end{document}